\documentclass[twocolumn,showpacs,preprintnumbers,amsmath,amssymb]{revtex4}

\usepackage{color,graphicx}
\usepackage{dcolumn}
\usepackage{bm}

\begin{document}
\title{Optimization in Gradient Networks}

\author{Natali Gulbahce}
\email[E-mail address:]{gulbahce@lanl.gov}
\affiliation{Theoretical Division and Center for Nonlinear Studies, Los Alamos 
National Laboratory, MS B284, Los Alamos, NM 87545, USA} 

\date{\today}

\begin{abstract}
Gradient networks can be used to model the dominant structure of complex
networks. Previous works have focused on random gradient networks.  Here we
study gradient networks that minimize jamming on substrate networks with
scale-free and Erd\H{o}s-R\'enyi structure.  We introduce structural 
correlations and strongly reduce congestion occurring on the network by using 
a Monte Carlo optimization scheme. This optimization alters the degree 
distribution and other structural properties of the resulting gradient
networks. These results are expected to be relevant for transport and other
dynamical processes in real network systems.  
\end{abstract} 

\maketitle

{\bf \noindent
Complex networks have been shown to offer a powerful framework for the  study
of dynamical processes in complex systems \cite{general_1,general_2}. A
fundamental question in this context is to identify how the structure of
dominant connections influences the dynamical properties of the entire system.
Gradient networks have been introduced in references \cite{zoltan,zoltan2}
precisely to characterize the structure governing transport in complex
networks.  Remarkably, similar structures have been identified recently in the
optimization of synchronizability in oscillator networks
\cite{grad_application1,grad_application2}.  Here, we consider the problem of
optimization of transport and show that by using a Monte Carlo approach congestion
can be reduced in complex networks. We discuss the structural correlations that 
emerge with optimization and its implications for real systems.
}

\section{Introduction}
\label{int}

The efficiency of transport systems has been of interest in various
fields, including physics, biology and engineering. In transport processes, the 
item being transported usually follows the steepest descent of the
underlying surface, e.g., water flowing down the slopes of a mountain. Flow in
networks has been modeled by Toroczkai et al. \cite{zoltan,zoltan2} by the
introduction of local gradients on a substrate network. The gradient 
network defined on this network has provided significant insights into the 
dominant structures that provide transport efficiency. They have considered a 
fixed network of $N$ nodes with a scalar potential, $V_i$ 
at each node $i$. The gradient $\nabla V_i$ of the potential at each node 
$i$ is a directed edge which points from $i$ to the neighbor with the minimum 
potential among all the neighbors of $i$.


Toroczkai et al have shown several properties of gradient networks which we
briefly summarize. An interesting topological property of the gradient network
is that its in-degree distribution is scale-free for both scale-free (SF) and
Erd\H{o}s-R\'enyi (ER) substrate networks~\cite{general_2}. A gradient network
with a non-degenerate potential distribution, is a group of trees, hence no
loops exist in this network other than self-loops. Only this property makes the
gradient networks very common in seemingly unrelated problems, i.e.,
synchronization in oscillatory networks
\cite{grad_application1,grad_application2}.  The relationship between network
topology and congestion has also been investigated \cite{zoltan,zoltan2} by
introducing a measure of congestion, the jamming coefficient. This measure
involves the ratio of the number of nodes that receive at least one gradient
link, $N_{\rm receive}$ and the number of nodes that send a link.  By
definition, every node sends one out link, therefore the number of senders in
the network, $N_{send}=N$. The jamming coefficient is

\begin{equation}
\label{eq:jam} 
J=1-\left \langle \!\left \langle N_{\rm receive}/N_{\rm send} \right \rangle_V\right \rangle_{\rm network}.
\end{equation}

The operations $\langle \ldots \rangle_V$ and $\langle \ldots \rangle _{\rm
network}$ denote the statistical averaging over local potentials and networks
respectively. Maximal congestion occurs at $J=1$, and no congestion occurs
when every link receives a gradient link, corresponding to $J=0$.  It was also
found that $J$ is independent of number of nodes $N$ for scale free substrate
network, and these networks are not prone to maximal jamming.  The jamming
coefficient was extensively studied by Park et al.~\cite{park} where they compared
it for ER and SF networks with the same average degree, $\langle k \rangle$, for 
$2 < \langle k \rangle < 200$. With randomly assigned potentials on each node, 
below $\langle k \rangle \approx 10$, they found that ER networks are less 
congested than SF networks.


Here we introduce a Monte Carlo optimization scheme that reduces jamming
significantly and introduces structural correlations into the system that are 
not built in. The remainder of the paper is organized as follows. In
Section~\ref{sec:opt} we introduce the algorithm for optimizing jamming
coefficient that is initially calculated from randomly assigned scalar values
at each node. We compare the optimized jamming values of a scale-free network
and an Erd\H{o}s-R\'enyi network for various values of average degrees. In
Section~\ref{sec:structure} we investigate the structure of the optimal
gradient network. In particular we study the degree distribution and the
correlations between the degree and the potential of each node. Finally in
Section~\ref{sec:conc} we discuss the implications and the possible extensions
of optimal gradient networks.

\section{Optimization of Jamming}
\label{sec:opt}


Previous work ~\cite{zoltan,zoltan2} has focused on random gradient networks
where the potential on each node has a randomly assigned value.  More
generally, the potentials can be a dynamic quantity evolving in time due to
perturbations, sources and sinks internal and external to the system of
interest.  Alternatively, the potentials can evolve to become correlated to the 
network properties such as its degree distribution.  For example, consider 
the networks of routers where every router has a capacity. If a router is 
central and highly connected, it usually has a higher capacity in order to 
handle the traffic en-route effectively. Recently, a congestion aware routing 
algorithm has been introduced where the transport on the network of routers 
is driven by congestion-gradients \cite{danila}.

Here we develop a Monte Carlo algorithm to achieve two goals: reduce jamming in
the network, and observe the emerging optimal correlation between in-degree and
potential of each node. For a given network and potential distribution, we
redefine J in Eq.~\ref{eq:jam} as $J = 1 - N_{\rm receive}/N$ where 
$N_{\rm send} = N$ by definition. The initial
potential distribution is chosen from a Gaussian distribution, and at each
iteration the potential of a random node is modified such that global
congestion is reduced. We use a Metropolis algorithm \cite{barkema} with the
following steps:

\begin{enumerate}
\item Pick a node, $i$ at random.
\item Vary $V_i$ by $\delta V$, i.e., ${V_i}_{\rm new} = V_i + \delta V$ where $\delta V$ is a 
Gaussian random variable with variance $\sigma^2=1$.
\item Recalculate $J$ with ${V_i}_{\rm new}$.
\item Accept ${V_i}_{\rm new}$ with probability $p \sim \exp[-\Delta J /T]$ where 
$\Delta J = J_{\rm new} - J_{\rm old}$.
\item Go to step 1, and repeat.
\end{enumerate}
The fictitious temperature, $T$ is chosen to adjust the acceptance ratio to
about 40\%.  We perform the optimization procedure until $J(t)$ equilibrates, 
i.e., at large $t$ the time autocorrelation function of $J(t)$ \cite{gould}, 
\begin{equation}
C_J(t) = \frac{[\langle J(t)J(0) \rangle - \langle J \rangle ^2]}{ [\langle J^2 \rangle - \langle J \rangle^2]}, 
\end{equation}
goes to zero.

\begin{figure}[h]
\includegraphics[width=8.5cm]{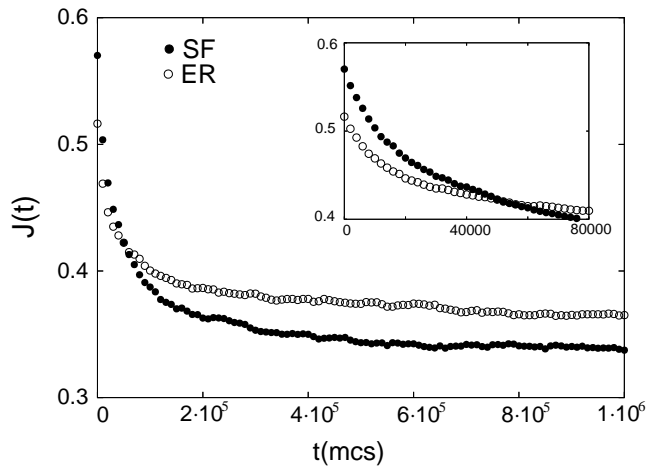}
\caption{Optimized jamming, $J(t)$, as a function of time in Monte Carlo steps
(mcs) for SF ($\bullet$) and ER ($\circ$) networks with the same average degree, $<k>=4$ and $N=10000$. 
As shown in the inset, SF network with random potential distribution 
has a higher jamming coefficient at $t=0$ compared to the ER network. After 
optimization is completed, however, optimal SF network has a lower jamming coefficient.}
\label{fig:fig1}
\end{figure}

Using the described optimization algorithm, we can significantly reduce the
jamming coefficient in both SF and ER networks.  Following previous work
\cite{zoltan,park} we choose as the SF network a Barab\'asi-Albert
model~\cite{general_2,science}. For this network the average connectivity is
$\langle k \rangle = 2 m$ where each node has at minimum $m$ links. The 
network size throughout the paper is chosen to be $N=10000$. For the
ER networks, $\langle k \rangle = pN$ where $p$ is the probability of having
a link between any pair of nodes. We use the same $\langle k \rangle$ when
comparing the two types of network by adjusting $m$ and $p$.  The evolution of
jamming coefficient during optimization is shown in Fig.~\ref{fig:fig1} as a
function of algorithmic time in units of Monte Carlo steps (mcs) for $\langle k
\rangle = 4$. At $t=0$ (see also inset of Fig.~\ref{fig:fig1}), the SF network
has a higher jamming coefficient, but after roughly 50000 mcs, the SF network
becomes less congested compared to the ER network. The initial and final
jamming coefficients for the two networks are ${J_i}_{SF} =
0.57$, ${J_i}_{ER} = 0.52$, ${J_f}_{SF} = 0.31$, and ${J_f}_{ER} = 0.36$,
respectively.

\begin{figure}[h] 
\includegraphics[width=8.5cm]{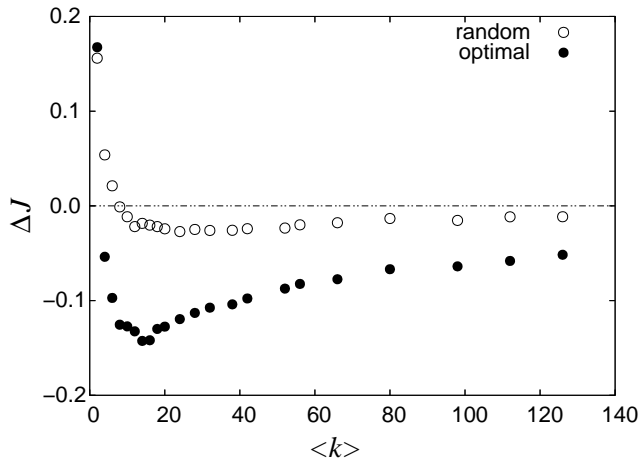}
\caption{Difference between the jamming values of SF and ER networks,
$\Delta J = J_{SF} - J_{ER}$ for various $\langle k \rangle $ for random 
($\circ$) and optimal ($\bullet$) gradient 
networks. $\Delta J < 0$ for all simulated $\langle k\rangle > 2$ indicating 
that optimal SF networks have a lower jamming coefficient than optimal ER 
networks.\label{fig:fig2}} 
\end{figure}

We define $\Delta J_r$ and $\Delta J_o$, the difference in jamming between SF
and ER networks for random and optimal networks respectively.  For a given
network with $\langle k \rangle$, $\Delta J_r$ is calculated initially at $t=0$
and $\Delta J_o$ at $t_{\rm opt}$ after optimization is completed, for $2\leq
\langle k \rangle \leq 126$. As shown in Fig.~\ref{fig:fig2}, $\Delta J_o<0$
for $\langle k \rangle > 2$ whereas $\Delta J_r>0$ indicating that SF networks
have a lower jamming coefficient after optimization, a result significantly
different than those for random gradient networks \cite{park}.

\section{Structural Properties of Optimal Networks}
\label{sec:structure}


As shown in Fig.~\ref{fig:fig1} congestion can be reduced in gradient networks by 
varying the potentials at each node. It is reasonable to expect that the obtained 
optimal potentials may also alter the structure of the gradient network.
Next, we analyze the structural properties of optimal gradient networks such 
as the degree distribution. Previously, the random gradient network of an ER 
substrate network was shown numerically and analytically \cite{zoltan,zoltan2} 
to have an in-degree distribution of $R(l)\sim l^{-1}$. However, the SF substrate 
network with degree distribution $P(k) \sim k^{-3}$, was shown to have a gradient 
degree distribution of $R(l) \sim l^{-3}$. 

\begin{figure}[h]
\includegraphics[width=8.5cm]{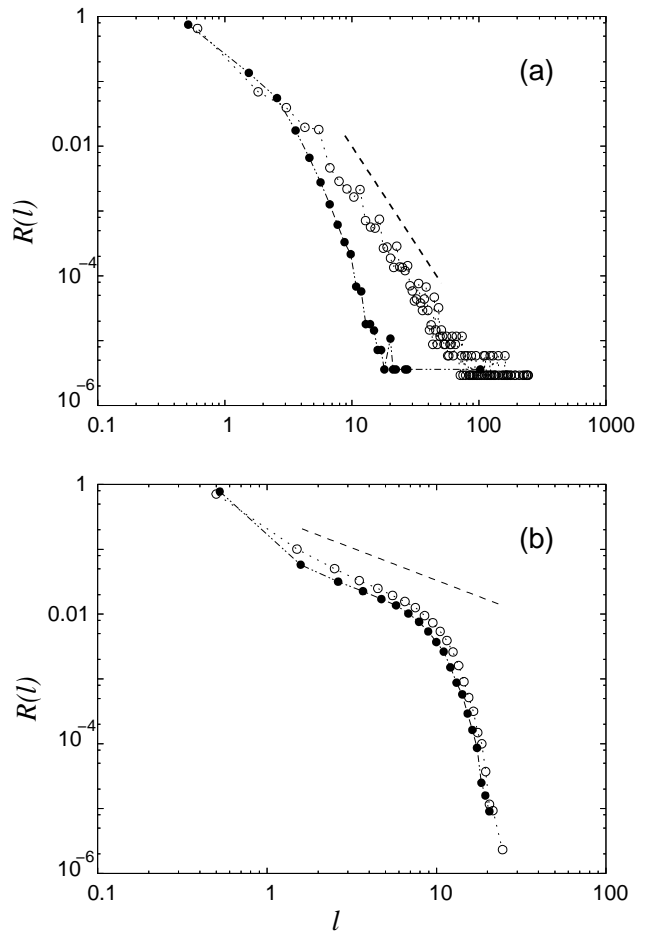}
\caption{In-degree distributions, $R(l)$, for gradient networks of (a) SF and (b) ER 
networks with $m=3$ and $p=0.001$ respectively before ($\circ$) and after 
($\bullet$) optimization of jamming. The dashed lined indicates the 
scaling of $R(l) \sim l^\alpha$ from Ref.~\cite{zoltan, zoltan2}, where $\alpha=-3$ and 
$-1$ for random SF and ER networks respectively. The scaling exponent $\alpha$ is not a 
good fit for optimal SF data but a good fit for optimal ER network. \label{fig:fig3}} 
\end{figure}

The degree distribution of the random and optimal ER and SF networks are shown 
for $p=0.001$ ($\langle k \rangle =10$) and $m=3$ ($\langle k \rangle =6$) 
respectively in Figure~\ref{fig:fig3} along with the 
expected scaling exponents for the random gradient networks of $-1$ and $-3$.
The statistical averaging is obtained over 100 networks. For ER networks, the 
scaling region extends with higher average connectivity, however we chose to 
use a small $\langle k \rangle =10$ for which optimization is more efficient.
The optimization is performed for 1 million Monte Carlo steps. The jamming
values initially are 0.62, 0.71 and after optimization reduce to 0.36, 0.60 for 
SF and ER respectively. The in-degree distribution, $R(l)$, for the SF network varies 
significantly with the optimization. The cut-off degree is reduced an order of 
magnitude (from roughly 100 to 10) compared to the random one, and the scaling is now 
steeper. On the other hand, the ER network does not show a major change in the degree
distribution.  

\begin{figure}[h] 
\includegraphics[width=8.5cm]{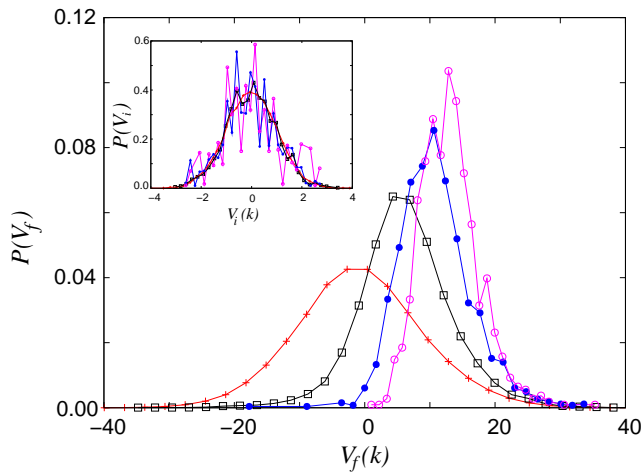} 
\caption{Probability distribution of the potentials of nodes with degree $k$
for SF network with $m = 3$ before (inset) and after optimization.  The degrees
are binned to indicate the correlation between degree of a node and its
potential: $3 < k \leq 10$ ({\color{red}$+$}), $10 < k\leq 30$ ($\square$), $30
< k \leq 50$ ({\color{blue} $\bullet$}), $50 < k\leq 100$ ({\color{magenta}
$\circ$}).  Before the optimization $P(V_i)$ do not show any correlations with
$k$, however after the optimization, nodes with high degree, get the large
values of potential which facilitates reduced jamming. \label{fig:fig4}}
\end{figure}

Next, we analyze the probability distribution of potentials for nodes in the
substrate networks with degree $k$, before and after the optimization to
observe any degree-potential correlations. The results are shown in
Fig.~\ref{fig:fig4} for the SF network ($m=3$).  To reduce noise in the data
especially for large values of $k$, the degrees are binned into four groups: $3
< k \leq 10$, $10 < k\leq 30$, $30 < k \leq 50$, $50 < k\leq 100$. Before
optimization with initial random Gaussian potentials within range [-4, 4] ,
each set has the same probability distribution $P(V_i)$ as shown in the inset
of Fig~\ref{fig:fig4}.  This behavior is expected as no correlation was built
in between the degree of a node and its potential. After the optimization
however, the range of the potentials has broadened significantly, and the nodes 
with high degree have accumulated very large potentials. 

With the improved jamming coefficient, it is natural to expect some correlation
to emerge between the potential of the node and its degree. If the node has a
large degree and a small potential, this node will be preferred by most of its
neighbors for sending an out link, and thus this will contribute to higher
jamming. However if the potential is large, the neighboring nodes will not
prefer this highly connected node and thus not increase the jamming. This
intuitive observation implies the possibility of obtaining a reduced congestion
by starting with potentials that are inversely correlated with the degree of
each node on the substrate network.  We tested this case for the SF network ($m=3$) 
with a correlated potential distribution, $V_i = r / k_i$ at node $i$ with 
degree $k_i$ where $r$ is a random number chosen from a uniform distribution. This 
assignment with correlations built-in did not make the jamming coefficient lower. On 
the contrary it was higher, $J=0.77$, than the value without degree correlation, $J=0.62$.

An interesting observation that Fig.~\ref{fig:fig4} provides is that nodes with
small degree carry potentials distributed over a large range $[-40,40]$. For
example, nodes with degrees, $3<k<10$ have a roughly Gaussian potential
distribution. For higher $k$, the distribution narrows down and shifts toward
larger values. For nodes with $50<k<100$, all nodes have large potentials,
within range $[0,40]$. This observation might explain why in the test case the
jamming was actually higher when the degree was correlated with the potential.
In that case, we only assigned low potentials to low degree nodes which still
yields congestion much higher than the optimal one.  An analytical formulation
that distributes the potentials according to its degree mimicking the
transitive behavior in Fig.~\ref{fig:fig4} seems possible, but is beyond the
scope of this paper. 

\section{Conclusion}
\label{sec:conc}

We have introduced a Monte Carlo method to optimize congestion in random 
gradient networks. Previously the potentials have been assigned randomly and 
was shown that ER networks had lower jamming coefficient below 
$\langle k \rangle =10$ than SF networks with the same connectivity \cite{park}. This was
puzzling as the connectivity commonly observed in natural and man-made
networks~\cite{general_2} is usually in this range, but tends to be scale-free, 
and in scale-free networks jamming is independent of $N$. With the Monte Carlo 
based optimization scheme we optimized jamming by varying the potentials so 
that optimal congestion was achieved.  We found that optimal SF networks 
have lower congestion factor for $\langle k \rangle > 2$.
This reduced congestion is the result of a complex correlation between the
degree and the potential of a node. We found that nodes with large degrees in the
substrate network get large positive values whereas nodes with small degrees get a
Gaussian like distribution of potentials.

Throughout the paper we have used the definition of jamming introduced in Ref.~\cite{zoltan}
for a substrate network with unweighted links. A natural extension of this work for 
generality is to assign weights to links and redefine the jamming coefficient accordingly. 
A possible definition is $ J = \big(\sum_i \big[ \sum_j w_{ij} - c_i \big ] \big ) / \sum_i c_i$
where $i=1 \cdots N$, $j$ is the number of neighboring links node $i$ has, $c_i$
is the capacity of node $i$, $w_{ij}$ is the weight of the incoming link from j to i, 
and the operation $[x] = 0$ if $x<0$. If the weights and capacity of all nodes
are 1, then this definition of $J$ reduces to the one in Eq.~\ref{eq:jam} without the averaging.  
With this definition and the optimization method introduced in the paper, it is possible
to study real world networks and get insights to the dominant structures of transport
for these systems.

\begin{acknowledgments}
The author thanks Adilson E. Motter for useful discussions and suggestions,
Gregory Johnson and Frank Alexander for the careful reading of the manuscript.
This work was carried out under the auspices of the National Nuclear Security
Administration of the U.S. Department of Energy at Los Alamos National
Laboratory under Contract No.DE-AC52-06NA25396 and supported by the DOE Office
of Science ASCR Program in Applied Mathematics Research.

\end{acknowledgments}


\end{document}